\documentclass[fleqn,usenatbib]{mnras}

\usepackage{newtxtext,newtxmath}

\usepackage[T1]{fontenc}

\DeclareRobustCommand{\VAN}[3]{#2}
\let\VANthebibliography\thebibliography
\def\thebibliography{\DeclareRobustCommand{\VAN}[3]{##3}\VANthebibliography}

\usepackage{graphicx}	
\usepackage{amsmath}	
\usepackage{color}

\newcommand{\Zsun}{Z_{\odot}}
\newcommand{\Msun}{M_{\odot}}

\newcommand{\Mdisk}{M_{\rm{disk}}}
\newcommand{\Mstar}{M_{\rm{star}}}

\newcommand{\revred}{\textcolor{black}}

\newcommand{\magB}{\mathbf{B}}

\newcommand{\ms}{~{\rm m} ~{\rm s}^{-1} }

\newcommand{\mum}{{\rm \mu} {\rm m} }

\newcommand{\psec}{\rm{s}^{-1}}
\newcommand{\gcc}{{\rm g\ cm}^{-3}}

\title[Formation and early evolution of disks in various space environments]{Cosmic-ray ionization rate versus Dust fraction :\\Which plays a crucial role in the early evolution of the circumstellar disk?}

\author[Y. Kobayashi et al.]{
Yudai Kobayashi,$^{1}$\thanks{E-mail: k7983964@kadai.jp}
Daisuke Takaishi$^{1}$
 and Yusuke Tsukamoto$^{1}$
\\
$^{1}$Graduate School of Science and Engineering, Kagoshima University, Kagoshima 890-0065, Japan
}

\date{Accepted XXX. Received YYY; in original form ZZZ}

\pubyear{2022}

\begin{document}
\label{firstpage}
\pagerange{\pageref{firstpage}--\pageref{lastpage}}
\maketitle

\begin{abstract}
We study the formation and early evolution of young stellar objects (YSOs) using three-dimensional non-ideal magnetohydrodynamic (MHD) simulations to investigate the effect of cosmic ray ionization rate and dust fraction (or amount of dust grains) on circumstellar disk formation.
Our simulations show that a higher cosmic ray ionization rate and a lower dust fraction lead to (i) a smaller magnetic resistivity of ambipolar diffusion, (ii) a smaller disk size and mass, and (iii) an earlier timing of outflow formation and a greater angular momentum of the outflow.
In particular, at a high cosmic ray ionization rate, the disks formed early in the simulation are dispersed by magnetic braking on a time scale of about $10^4$ years.
Our results suggest that the cosmic ray ionization rate has a particularly large impact on the formation and evolution of disks, while the impact of the dust fraction is not significant.

\end{abstract}




\section{Introduction}
Molecular cloud cores, which are the parent bodies of protostars and protostellar disks, are strongly magnetized \citep{2008ApJ...680..457T,2010ApJ...725..466C,2012ARA&A..50...29C}.
For example, molecular cloud cores typically have a mass-to-magnetic-flux ratio of 2.0, as revealed by the Zeeman effect \citep{2008ApJ...680..457T}.


Such a strong magnetic field removes angular momentum from the central region (known as magnetic braking) and suppresses disk formation. 
For example, \citet{1994ApJ...432..720B} shows that magnetic braking causes an exponential decrease in angular velocity.
The molecular cloud core is composed of weakly ionized gases \citep{1990MNRAS.243..103U,2002ApJ...573..199N,2012ApJ...759...35I}.
Therefore, non-ideal MHD effects (the Ohmic dissipation, ambipolar diffusion, and Hall effect) that determine the coupling rate between the magnetic field and gas are important.
\citet{2011PASJ...63..555M}, \citet{2015MNRAS.452..278T} and \citet{2013ApJ...763....6T}
showed that the Ohmic dissipation enables the formation of protoplanetary disks as large as 1 AU at the protostar formation epoch by eliminating the coupling between the magnetic field and gas in the first core \citep{1969MNRAS.145..271L}.
The ambipolar diffusion decouples the magnetic field and gas even at lower densities than the Ohmic dissipation, removing magnetic fluxes from the disk and suppressing the magnetic braking over a wider density range \citep{1991ApJ...371..296M, 2012A&A...541A..35D, 2016A&A...587A..32M, 2016MNRAS.457.1037W, 2021MNRAS.507.2354W, 2016A&A...592A..18M, 2016MNRAS.460.2050Z, 2018MNRAS.473.4868Z, 2015ApJ...810L..26T, 2017ApJ...838..151T, 2015ApJ...801..117T}.
Therefore, ambipolar diffusion is considered to play a major role in the evolution of magnetic fluxes at standard metal (or dust) fractions and cosmic ray ionization rates.

According to observations, different galaxies and star-forming regions have different cosmic ray ionization rates and metallicities.
\citet{2018MNRAS.480.5167P}, \citet{1998ApJ...503..689W} and \citet{1998ApJ...499..234C} showed that the cosmic ray ionization rate decreases with increasing column density in the molecular cloud.
Furthermore, \citet{2009A&A...506.1137C} has found that the metallicity is up to 10 times lower in the Large Magellanic Cloud, Small Magellanic Cloud, and M33 galaxies.

The cosmic ray ionization and dust grains (the amount of which is proportional to the metallicity) act as sources and sinks of charged particles, respectively, and affect the ionization degree of the gas.
Therefore, the ionization degree will be high in an environment with high cosmic ray ionization and/or low metallicity.
Since non-ideal MHD effects are weaker in gases with a high ionization degree, it is expected that the magnetic braking will be more efficient and that disk formation and its growth will be suppressed.
This may be related to the observation by
\citet{2010ApJ...723L.113Y} that the lifetime of circumstellar disks is shortened in low metallicity environments.

On the other hand, theory shows that in highly ionized environments, the gas is well coupled to the magnetic field and the angular momentum is removed by magnetic braking, thus preventing the formation of disks.
For example, \citet{2018MNRAS.476.2063W,2018MNRAS.481.2450W} and \citet{2020A&A...639A..86K} were inspired by observations of differences in cosmic ray intensity in different star-forming regions to study the effect of cosmic ray ionization rates on disk formation.
They found that disks formed in environments with low cosmic ray ionization rates, while disks did not form in environments with high cosmic ray ionization rates, a dramatic change.
However, because the effect of metallicity (or dust fraction) was not covered in their study, the effects of metallicity on the evolution of YSOs, including the evolution of disks, have not yet been fully clarified.

As the first step of a series of future studies, we investigate the effect of dust fraction (or amount of dust grains) and cosmic ray ionization rate on the formation and evolution of YSOs using 3D non-ideal MHD simulations.

\section{Numerical method and initial condition}

\subsection{Numerical Method}
We solved non-ideal magnetohydrodynamics (MHD) equations,

\begin{eqnarray}                                                                
\frac{D \mathbf{v}}{D t}&=&-\frac{1}{\rho}\left\{ \nabla \left( P+\frac{1}{2}|\magB|^2 \right) - \nabla \cdot (\mathbf{ B B})\right\} \nonumber \\
- \nabla \Phi,  \\                                                              
\frac{D {\mathbf B}}{D t}  &=& \left( {\mathbf B} \cdot \nabla \mathbf{v} -  {\mathbf B} (\nabla \cdot \mathbf{v})\right)    \nonumber \\                      
&-& \nabla \times \left\{ \eta_O (\nabla \times \mathbf B)  \right.    \nonumber \\                                                                            
&-& \left. \eta_{\rm A} ((\nabla \times \mathbf B) \times \mathbf {\hat {B}}) \times \mathbf {\hat {B}}\right\},  \\                                                 
\nabla^2 \Phi&=&4 \pi G \rho,\\                                                 
P &=& P(\rho)= c_{\rm s, iso}^2 \rho \left\{ 1+\left(\frac{\rho}{\rho_{\rm crit}}\right)^{2/3} \right\},                                                  
\end{eqnarray}
where $\rho$ is the gas density, $P$ is gas pressure, $\magB$ is magnetic field, $\Phi$ is the gravitational potential, and $G$ is a gravitational constant.
$\hat{\magB}$ is defined as $\hat{\magB}\equiv\magB/|\magB|$.
$\eta_{\rm O}$ and $\eta_{\rm A}$ are the resistivities for the Ohmic dissipation and ambipolar diffusion, respectively. 
We used a barotropic equation of state in which gas pressure only depends on density.
$c_{\rm s, iso}=190 \ms$ is isothermal sound velocity at $10$ K.
We used a  critical density of $\rho_{\rm crit}=4\times 10^{-14} \gcc$ above which gas behaves adiabatically.
In this study, we ignored the Hall effect because the calculation cost was enormous.

We used the smoothed particle magneto-hydrodynamics (SPMHD) method to solve the equations
\citep{2011MNRAS.418.1668I,2013ASPC..474..239I}.
See \citet{2020ApJ...896..158T} for details on numerical calculations.

To avoid small time-stepping, we employed the sink particle technique \citep{1995MNRAS.277..362B}.
The sink particle is dynamically introduced when the density exceeds
$\rho_{\rm sink}=4 \times 10^{-13} \gcc$.
The sink particle absorbs SPH particles
with $\rho>\rho_{\rm sink}$ within $r<r_{\rm sink}=1$ AU, and the mass and linear momentum of SPH particles are added to those of sink particles.

\subsection{Initial conditions}
We used the density-enhanced Bonnor-Ebert sphere surrounded by medium with a steep density profile of $\rho \propto r^{-4}$ as the initial density profile, which is the same as our previous studies \citep{2021ApJ...920L..35T},
\begin{eqnarray}                                                                             
  \rho(r)=\begin{cases}
  \rho_0 \rho_{\rm BE}(r/a) ~{\rm for} ~ r < R_c \\
  \rho_0 \rho_{\rm BE}(R_c/a)(\frac{r}{R_c})^{-4} ~{\rm for} ~ R_c < r < 10 R_c,
    \end{cases}
\end{eqnarray}
and
\begin{eqnarray}
    a=c_{\rm s, iso} \left( \frac{f}{4 \pi G \rho_0} \right)^{1/2},
\end{eqnarray}
where $\rho_{\rm BE}$ is the non-dimensional density profile of the critical Bonnor-Ebert sphere, $f$ is a numerical factor related to the strength of gravity, and $R_c=6.45 a$ is the radius of the cloud core.
$f=1$ corresponds to the critical Bonnor-Ebert sphere, and the core with $f>1$ is gravitationally unstable.
A Bonnor-Ebert sphere is determined by specifying the central density $\rho_0$, the ratio of the central density to the density at $R_c$ $\rho_0/\rho(R_c)$, and $f$.
In this study, we used the values of $\rho_0=7.3\times 10^{-18} \gcc$, $\rho_0/\rho(R_c)=14$, and $f=2.1$. 
Then, the radius of the core is $R_c=4.8\times 10^3$ AU, and the enclosed mass within $R_c$ is $M_c=1 \Msun$.
The thermal energy of the central core (without surrounding medium) is denoted by the terms $E_{\rm therm}$ and $E_{\rm grav}$, respectively.
The $\alpha_{\rm therm}$ ($\equiv E_{\rm therm}/E_{\rm grav}$) is equal to $0.4$.
The steep envelope was used in order to place the outer boundary far from the cloud core's center.
With the steep profile, the total mass of the entire domain remains $\sim 2 M_c$.
For rotation of the cloud core, we used an angular velocity profile of $\Omega(d)=\frac{\Omega_0 }{\exp(10(d/(1.5 R_c))-1)+1}$ 
where $d=\sqrt{x^2+y^2}$ and $\Omega_0=2.3\times 10^{-13} {\rm s^{-1}}$.
$\Omega(d)$ is almost constant for $d<1.5 R_c$ and rapidly decreases for $d > 1.5 R_c$.
The rotating energy of the core, denoted by $E_{\rm rot}$, has a ratio to gravitational energy $\beta_{\rm rot}$($\equiv E_{\rm rot}/E_{\rm grav}$) equals $0.03$.

The magnetic field profile has only the $z$ component. The magnetic field strength and plasma $\beta$ at the center were $B_0=62 \mu G$ and $\beta=1.6 \times 10^1$, respectively. 
The mass-to-flux ratio of the core $\mu$ relative to the critical value was $\mu/\mu_{\rm crit}=(M_c/\Phi_{\rm mag})=3$. where $\Phi_{\rm mag}$ the magnetic flux of the core and $\mu_{\rm crit}=(0.53/3 \pi)(5/G)^{1/2}$.
We resolved 1 $\Msun$ with $3\times 10^6$ SPH particles. Thus, each particle has a mass of $m=3.3\times 10^{-7} \Msun$.

The model names and corresponding dust sizes are summarized in Table 1.

\subsection{Resistivity}
In this study, we choose dust fraction and cosmic ray ionization rate as parameters. 
Then, the resistivity was calculated by chemical reaction calculations with the parameters.
See \citet{2020ApJ...896..158T} for details on chemical reaction network calculation.

For the dust models, we considered a dust size distribution of $n(a)\propto a^{-3.5}$
with minimum and maximum dust sizes of $a_{\rm min}=0.005 \mum$ and $a_{\rm max}=0.25 \mum$ (MRN distribution). 
The dust internal density is fixed to be $\rho_d=2 \gcc$.
\revred{
The dust-to-gas mass ratio f is $10^{-2}$ or $10^{-3}$ and f=$10^{-2}$ is the solar dust-to-gas ratio.
}

The model names and corresponding dust fraction and cosmic ray ionization rate are summarized in Table 1.

\begin{table*}
	\centering
	\caption{Model name, dust fraction, the strength of cosmic ray, and dust model are given.}
	\label{tab:model_list}
	\begin{tabular}{lccr} 
		\hline
		Model name & \revred{f} & $\zeta_{\rm CR}[\psec]$ & dust model\\
		\hline
		\revred{model\_MRN\_f1e2\_zeta1e17}    & \revred{$10^{-2}$}  & $10^{-17}$ & MRN\\
		\revred{model\_MRN\_f1e3\_zeta1e17}   & \revred{$10^{-3}$} & $10^{-17}$ & MRN\\
		\revred{model\_MRN\_f1e2\_zeta1e16}    & \revred{$10^{-2}$}  & $10^{-16}$ & MRN\\
		\revred{model\_MRN\_f1e3\_zeta1e16}   & \revred{$10^{-3}$} & $10^{-16}$ & MRN\\
		\hline
	\end{tabular}
\end{table*}

\section{RESULTS}
\subsection{Impact on magnetic resistivity}
First, we examine how resistivity depends on cosmic ray ionization rate and dust fraction.

\begin{figure*}
    \includegraphics[width=0.3\linewidth,,angle=-90]{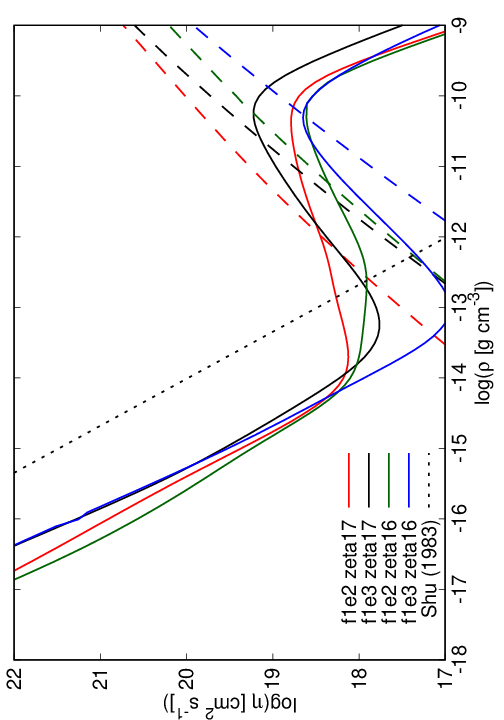}
    \includegraphics[width=0.3\linewidth,,angle=-90]{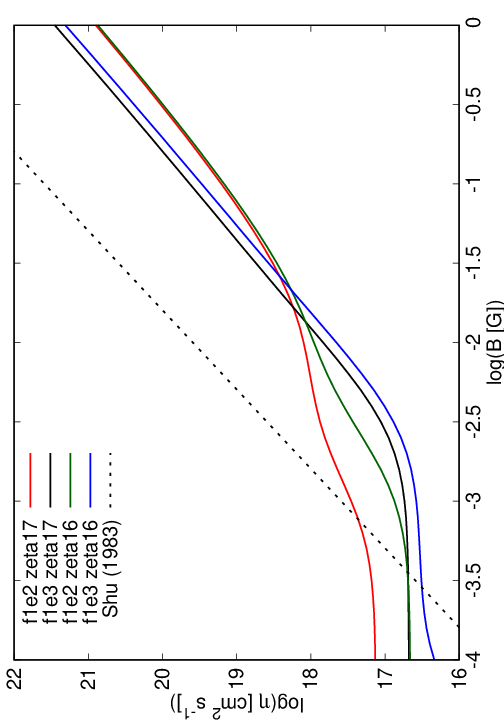}
    \caption{The left panel shows $\eta_{\rm A}$ \ (solid lines) and $\eta_O$ \ (dashed lines) as a function of density with a fixed magnetic field strength of $B=30 {\rm mG}$.
    The right panel shows $\eta_{\rm A}$ as a function of the magnetic field with a fixed density $\rho = 4\times10^{15} {\gcc}$.
    The dotted black line shows $\eta_{\rm A} = B^2 / (4\pi \gamma \rho C\sqrt{\rho})$ given by \citet{1983ApJ...273..202S}.We defined $\gamma = 3.5\times10^{13} {\rm{cm^3}\ g^{-1}\ s^{-1}}$ and $C=3.0\times10^{-16} {\gcc}$.}
    \label{fig:eta}
\end{figure*}

Figure \ref{fig:eta} shows the Ohmic resistivity $\eta_{\rm O}$ and ambipolar resistivity $\eta_{\rm A}$ with different cosmic ray ionization rates and dust fractions.
The left side of Figure \ref{fig:eta} shows $\eta_{\rm A}$ (solid line) and $\eta_{\rm O}$ (dashed line) as functions of density with a fixed magnetic field strength of $\magB = 3\times 10^{-2} \rm{G}$.
The black dotted line is $\eta_{\rm A} = B^2 / (4\pi \gamma \rho C \sqrt{\rho})$ \citep{1983ApJ...273..202S}, where we use $\gamma = 3.5 \times 10^{13} \rm{cm}^3\ \rm{g}^{-1}\ \rm{s}^{-1}$ and $C=3.0 \times 10^{-16} \rm{g}\ \rm{cm}^{-3}$.
In the low density region ($\lesssim 10^{-14} \gcc$), $\eta_{\rm A}$ decreases with increasing density.
In this density region, gases are ionized by a cosmic ray.
Thus, the fraction of the charged particle increases, and $\eta_{\rm A}$ decreases.
In the intermediate density region ($10^{-14}\ \gcc \lesssim \rho \lesssim 10^{-11}\ \gcc$), $\eta_{\rm A}$ increases with increasing density.
As the density increases, charged particles are absorbed by dust grains due to the increasing collision rate between dust grains and charged particles.
This absorption reduces the number of charged particles and lowers the conductivity.
Thus, $\eta_{\rm A}$ increases in the intermediate density.
In the high density region ($\gtrsim 10^{-11} \gcc$), $\eta_{\rm A}$ decreases with increasing density.

Note that the resistivity decreases with increasing cosmic ray intensity.
For example, for $\rho \sim 10^{-13} \gcc$, model\_MRN\_f1e2\_zeta1e16 has a twice lower $\eta_{\rm A}$ than model\_MRN\_f1e2\_zeta1e17, and model\_MRN\_f1e3\_zeta1e16 has an order of magnitude lower $\eta_{\rm A}$ than model\_MRN\_f1e3\_zeta1e17. 
Because cosmic rays ionize neutral gas, clouds have a higher ionization degree when the rays are stronger.
As a result, the clouds have higher conductivities and lower $\eta_{\rm A}$.

This indicates that $\eta_{\rm A}$ is affected by an increase in cosmic ray intensity and a decrease in dust fraction (solid line).
Also, $\eta_{\rm A}$  steepens with decreasing dust fraction in the density region ($10^{-14} \gcc \lesssim \rho \lesssim  10^{-10} \gcc$).

On the other hand, as cosmic rays and dust fraction increase, $\eta_{\rm O}$ decreases in the density region ($10^{-12}\ \gcc \lesssim \rho \lesssim 10^{-9}\ \gcc$).

The right panel of figure ~\ref{fig:eta} shows $\eta_{\rm A}$ as a function of magnetic field at $\rho=4\times10^{-15}\ \gcc$.
For all models, the dependence of $\eta_{\rm A}$ in a strong magnetic field ($\gtrsim 3 \times 10^{-2}\ {\rm{G}}$) is $\eta_{\rm A}\propto B^2$ as stated by \citep{1983ApJ...273..202S}.
In weak magnetic fields, on the other hand, the magnetic field dependence of $\eta_{\rm A}$ is not simple.
In all models, at least for $B<10^{-3}\ {\rm{G}}$, it is almost independent of the magnetic field, and the relation $\eta_{\rm A}\propto B^2$ is not necessarily true in the region $10^{-3}\ {\rm{G}} < B < 3\times 10^{-2}\ {\rm{G}}$.

\subsection{Time evolution of models in higher cosmic ray ionization and lower dust fraction environment}
In this section, we describe the time evolution of the disk and outflow of each model.

\begin{figure*}
    \includegraphics[width=1.2\linewidth,,angle=-90]{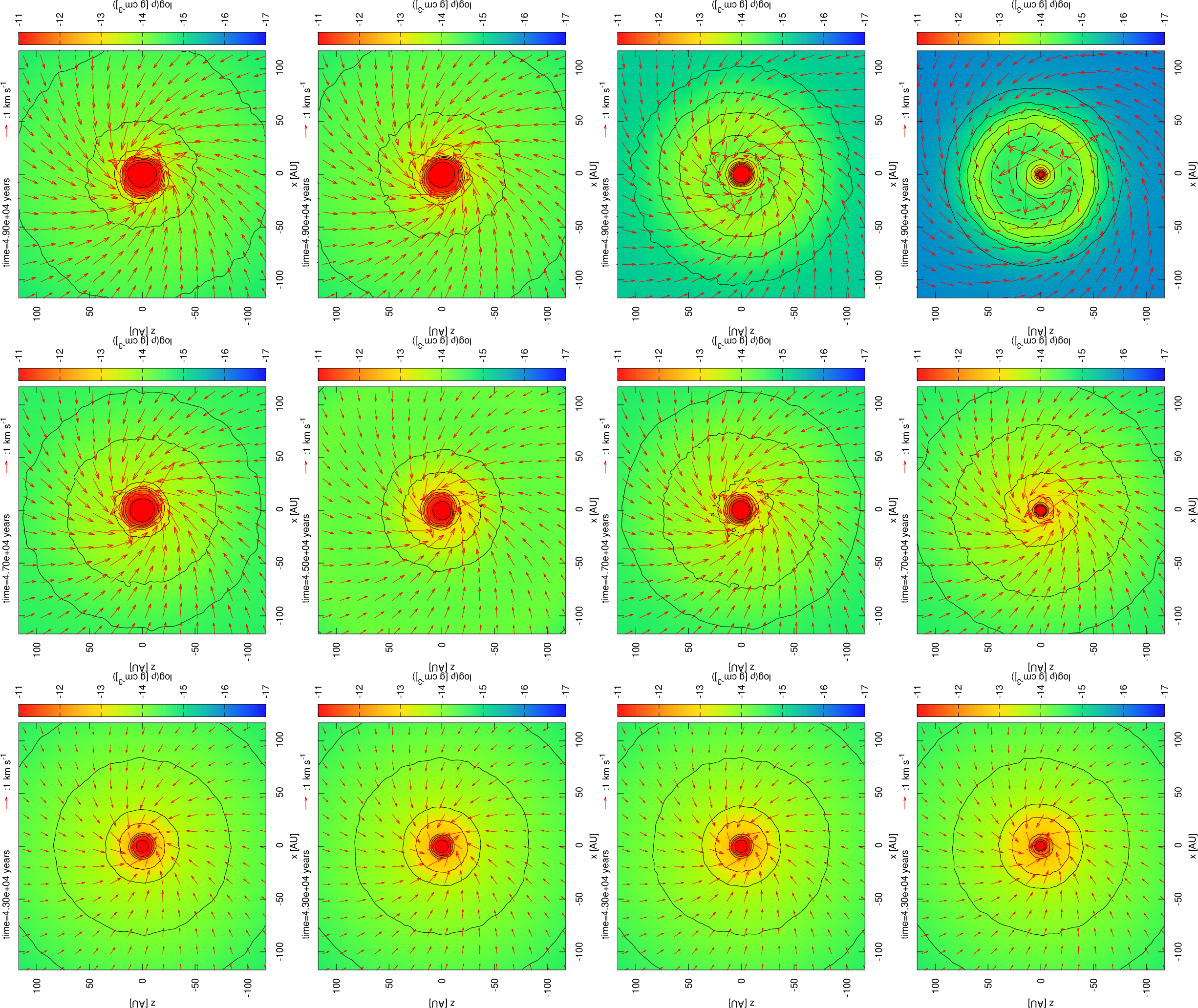}
    \caption{
    Density maps in two dimensions on the $x-y$ plane. for model\_MRN\_f1e2\_zeta1e17 \ (top), model\_MRN\_f1e3\_zeta1e17 \ (second row), model\_MRN\_f1e2\_zeta1e16 \ (third row), and model\_MRN\_f1e3\_zeta1e16 \ (bottom).
    The elapsed times are $t=4.3\times10^4$ yr \ (left), $t=4.7\times10^4$ yr \ (middle) and $t=4.9\times10^4$ yr \ (right).Red arrows show the velocity field.}
    \label{fig:xy_density_map}
\end{figure*}

Figure ~\ref{fig:xy_density_map} shows the density evolution on the $x$-$y$ plane in the 250 AU scale box of model\_MRN\_f1e2\_zeta1e17 (top), model\_MRN\_f1e3\_zeta1e17 (second row), model\_MRN\_f1e2\_zeta1e16 (third row), and model\_MRN\_f1e3\_zeta1e16 (bottom), at $t=4.3\times10^4$ yr (left column), $t=4.7\times10^4$ yr (middle column) and $t=4.9\times10^4$ yr (right column).
The circumstellar disk is the central high-density region ($\rho \gtrsim 10^{-14}\ \gcc$).
The disk is formed at $t=4.3\times10^4$ yr (left column) and is of similar size in all models.
At $t=4.7\times10^4$ yr (middle column), the disk in model\_MRN\_f1e3\_zeta1e16 is smaller than other disks.
At $t=4.9\times10^4$ yr (right column), model\_MRN\_f1e2\_zeta1e17 and model\_MRN\_f1e3\_zeta1e17 maintain the disk.
On the other hand, the disks shrink in model\_MRN\_f1e2\_zeta1e16 and model\_MRN\_f1e3\_zeta1e16.
A more quantitative analysis of the disk size evolution is presented in \S 3.3.

\begin{figure*}
    \includegraphics[width=1.2\linewidth,,angle=-90]{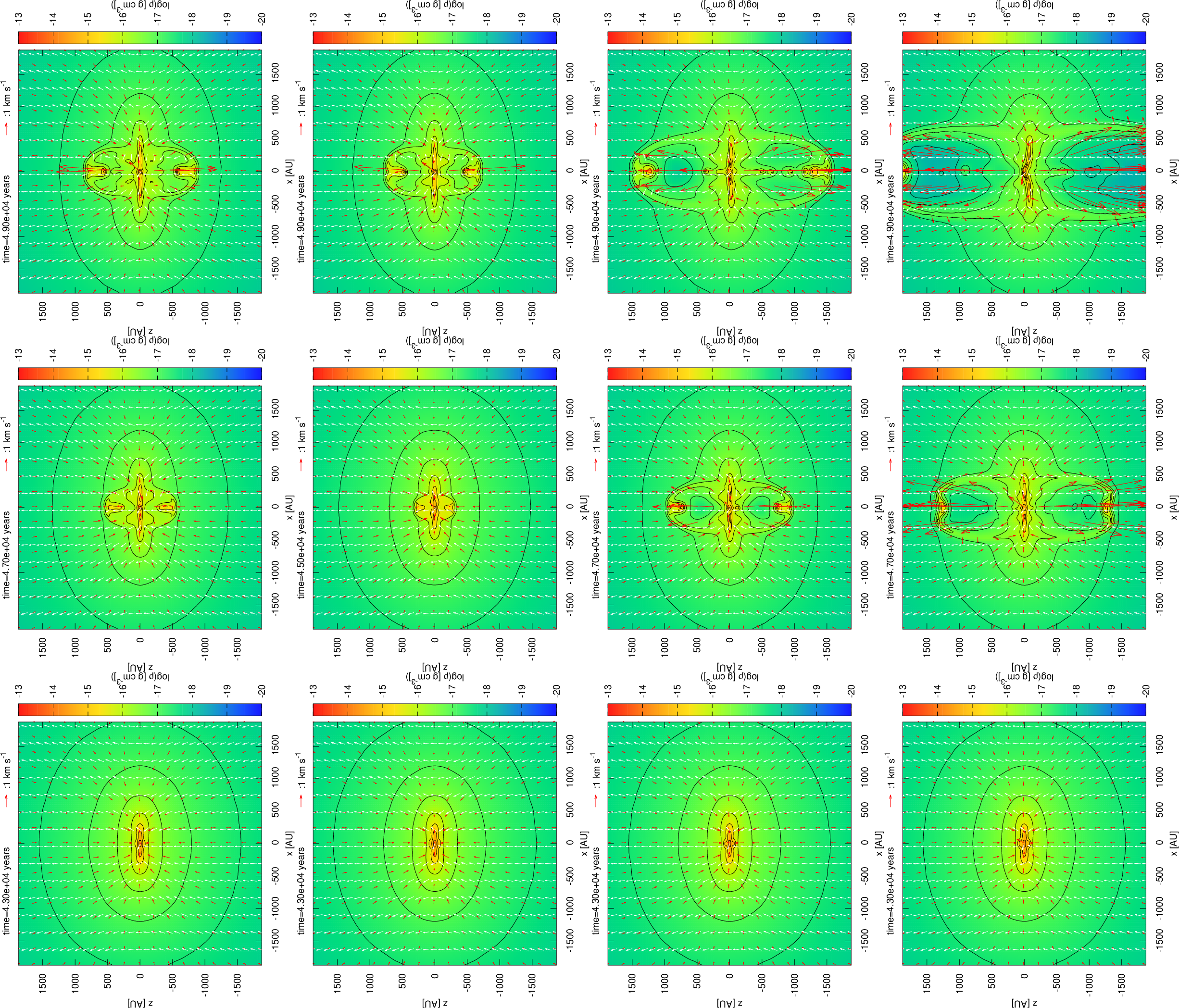}
    \caption{
    Density maps in two dimensions on the $x-z$ plane. for model\_MRN\_f1e2\_zeta1e17 \ (top), model\_MRN\_f1e3\_zeta1e17 \ (second row), model\_MRN\_f1e2\_zeta1e16 \ (third row), and model\_MRN\_f1e3\_zeta1e16 \ (bottom).
    The elapsed times are $t=4.3\times10^4$ yr \ (left), $t=4.7\times10^4$ yr \ (middle) and $t=4.9\times10^4$ yr \ (right).Red arrows show the velocity field, and white arrows show the direction of the magnetic field.}
    \label{fig:xz_density_map}
\end{figure*}

Figure ~\ref{fig:xz_density_map} shows the density evolution on the $x$-$z$ plane in the 4000 AU scale box of model\_MRN\_f1e2\_zeta1e17 (top), model\_MRN\_f1e3\_zeta1e17 (second row), model\_MRN\_f1e2\_zeta1e16 (third row), and model\_MRN\_f1e3\_zeta1e16 (bottom), for the same time period as figure ~\ref{fig:xy_density_map}.
All models formed the outflow in our simulation.
With high cosmic ray ionization (model\_MRN\_f1e2\_zeta1e16, model\_MRN\_f1e3\_zeta1e16), the outflow reaches further and the outflow velocity is faster than with lower cosmic ray (model\_MRN\_f1e2\_zeta1e17, model\_MRN\_f1e3\_zeta1e17).
While focusing on the difference in dust fractions, outflows have a large difference only when $\zeta=10^{-16} \rm{s^{-1}}$.
\S 3.5 contains a more quantitative analysis of the disk size evolution. 

\subsection{Time evolution of specific angular momentum and disk radius}

Here we examine the time evolution of the angular momentum and centrifugal radius in the central region to quantitatively discuss the disk size evolution after protostar formation.
The angular momentum of the disk $J(\rho_{\rm disk})$ is calculated by
\begin{align}
    J(\rho_{\rm disk})
    \equiv
    \left| \int_{\rho>\rho_{\rm disk}} \rho(\mathbf{r} \times \mathbf{v}) dV \right|.
\end{align}
For the density threshold of the disk, we use $\rho_{\rm disk}=1.0\times10^{-14} {\gcc}$.
At the first output after the sink particle is introduced, the centrifugal radius is calculated by
\begin{align}
    r_{\rm disk}
    \equiv r_{\rm cent}
    = \frac{\Bar{j}(\rho_{\rm disk})^2}{GM_{\rm star}}.
\end{align}
Here $\Bar{j}(\rho_{\rm disk}) = J(\rho_{\rm disk})/M(\rho_{\rm disk})$, where $M(\rho_{\rm disk})$ is the enclosed mass within the region $\rho>\rho_{\rm disk}$.
We regard this centrifugal radius as a disk radius.

The solid line in figure \ref{fig:t_J_comp} shows the time evolution of the centrifugal radius.
In low cosmic ray ionization models (red and black lines), the disk continues to grow with time after disk formation ($t \sim 4.3\times10^4$ yr) and until the end of the simulation ($t \sim 5.2\times10^4$ yr).
Even in models with high cosmic ray ionization (orange and yellow lines), the disk forms at $t \sim 4.3\times 10^4$ yr.
However, the disk size does not monotonically increase with time and eventually disappears at the end of the simulation.

The dashed line in figure \ref{fig:t_J_comp} shows the time evolution of the specific angular momentum of the disk.
The specific angular momentum increases after the disk forms in both the low and high cosmic ray ionization rate models.
In low models (red and black lines), it continues to increase with time, but in high models (orange and yellow lines), it declines after about 5,000 years.
Because gas is well-coupled with magnetic fields in the high cosmic-ray ionization rate models, the magnetic braking is stronger in these models than in the lower models.
As a result, the gas loses its angular momentum, and the disk is disappeared in the high cosmic ray ionization rate models.

\subsection{Time evolution of the mass of protostar and disk}
In this section, we investigate the time evolution of $\Mdisk$, $\Mstar$, and $\Mdisk + \Mstar$.
Disk mass is defined as the enclosed mass within $\rho>\rho_{\rm disk}=10^{-14} {\gcc}$.

\begin{figure}
\begin{minipage}[b]{\linewidth}
    \includegraphics[width=0.65\linewidth,angle=-90]{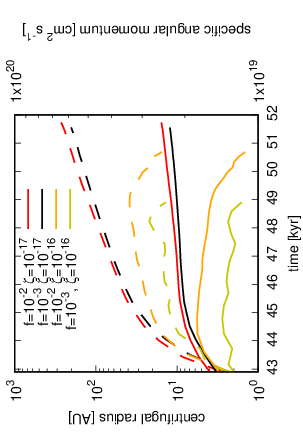}
    \caption{Time evolution of the centrifugal radius (solid lines; left axis) and specific angular momentum (dashed lines; right axis).
    The red, black, orange, and yellow lines show the results of model\_MRN\_f1e2\_zeta1e17, model\_MRN\_f1e3\_zeta1e17, model\_MRN\_f1e2\_zeta1e16, model\_MRN\_f1e3\_zeta1e16, respectively.}
    \label{fig:t_J_comp}
\end{minipage}\\
\begin{minipage}[b]{\linewidth}
    \includegraphics[width=0.65\linewidth,angle=-90]{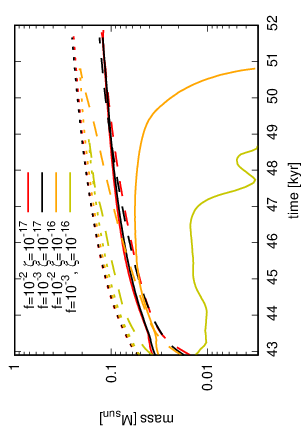}   
	\caption{Time evolution of the mass of disk $\Mdisk$ (solid lines), protostar $\Mstar$ (dashed lines), and $\Mdisk + \Mstar$ (dotted lines).
	The red, black, orange, and yellow lines show the results of model\_MRN\_f1e2\_zeta1e17, model\_MRN\_f1e3\_zeta1e17, model\_MRN\_f1e2\_zeta1e16, and model\_MRN\_f1e3\_zeta1e16, respectively.}
    \label{fig:t_mass_comp}
\end{minipage}
\end{figure}

Figure ~\ref{fig:t_mass_comp} shows the time evolution of disk mass $\Mdisk$ (solid), and protostar mass $\Mstar$ (dashed).
At $t=4.3\times10^4$ yr, the disk mass (solid) is comparable ($\Mdisk\sim0.2\Msun$) in model\_MRN\_f1e2\_zeta1e17, model\_MRN\_f1e3\_zeta1e17, and model\_MRN\_f1e2\_zeta1e16.
While model\_MRN\_f1e3\_zeta1e16 has a smaller disk mass.
With time, $\Mdisk$ increases in low cosmic ray ionization models (red and black lines), but in high cosmic ray ionization models (orange and yellow lines), $\Mdisk$ rapidly decreases at $t\sim4.7\times10^4$ yr and $t\sim4.9\times10^4$ yr, respectively.
This property is the same as disk radius and angular momentum (see figure ~\ref{fig:t_J_comp}).

On the other hand, the total masses of the protostars and disks (dotted) are almost the same for all models.
This indicates that the mass carried away by the outflow is sufficiently small and that the disappearance of the disk is due to mass accretion onto the central protostar.

\subsection{Time evolution of outflow}
In this section, we investigate the properties of outflow.
An outflow is defined as a region that satisfies $v_r = (\mathbf{v}\cdot\mathbf{r})/|\mathbf{r}| > 2c_{\rm{s,ios}}$ and $\rho < 10^{-14}\gcc$ where $c_{\rm{s,ios}}=0.19\rm{km}\ \rm{s}^{-1}$ is sound velocity at $T=10\rm{K}$.
This definition matches that in \citet{2020ApJ...896..158T}.
Outflows are formed in all models.

\begin{figure}
\begin{minipage}[b]{\linewidth}
    \centering
    \includegraphics[width=0.65\linewidth,,angle=-90]{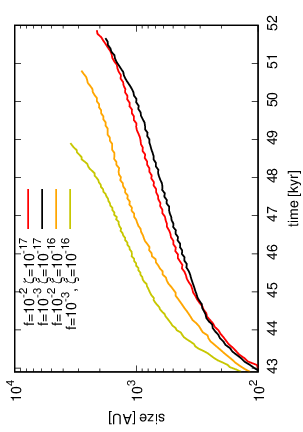}
    \caption{Time evolution of the size of outflow. The red, black, orange, and yellow lines show the results of model\_MRN\_f1e2\_zeta1e17, model\_MRN\_f1e3\_zeta1e17, model\_MRN\_f1e2\_zeta1e16, and model\_MRN\_f1e3\_zeta1e16, respectively.}
    \label{fig:outflow_size}
\end{minipage}
\begin{minipage}[b]{\linewidth}
    \centering
    \includegraphics[width=0.65\linewidth,,angle=-90]{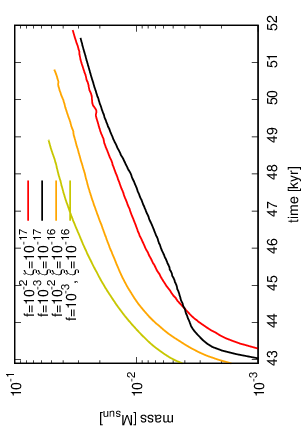}
    \caption{Time evolution of the mass of outflow. The red, black, orange, and yellow lines show the results of model\_MRN\_f1e2\_zeta1e17, model\_MRN\_f1e3\_zeta1e17, model\_MRN\_f1e2\_zeta1e16, and model\_MRN\_f1e3\_zeta1e16, respectively.}
    \label{fig:outflow_mass}
\end{minipage}
\begin{minipage}[b]{\linewidth}
    \centering
    \includegraphics[width=0.65\linewidth,,angle=-90]{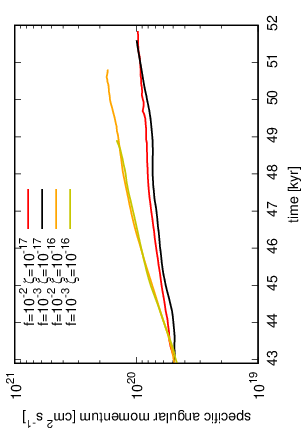}
    \caption{Time evolution of the specific angular momentum of outflow.The red, black, orange, and yellow lines show the results of model\_MRN\_f1e2\_zeta1e17, model\_MRN\_f1e3\_zeta1e17, model\_MRN\_f1e2\_zeta1e16, and model\_MRN\_f1e3\_zeta1e16, respectively.}
    \label{fig:outflow_J}
\end{minipage}
\end{figure}

Figure \ref{fig:outflow_size} shows the time evolution of the outflow size.
The outflow size is defined as the distance at which the particles in the outflow are farthest from the central protostar.
In our simulations, a strong outflow with velocity $v\ >\ 1\ \rm{km}\ \rm{s}^{-1}$ forms a few thousand years after protostar formation.
The size of the outflow increases monotonically.
In $t\sim9.0\times10^3$ yr after outflow formation with low cosmic ray ionization (red and black), the outflow size reaches $\sim 2000$ AU with both low and high dust fraction.
The average velocity of the outflow is thus 1.1$\ \rm{km}\ \rm{s}^{-1}$.
With high cosmic ray ionization, the outflow size reaches $ \sim 3000$ AU  at $t \sim8.0 \times10^3$ years (model\_MRN\_f1e2\_zeta1e16, orange line) and $t \sim6.0 \times10^3$ years (model\_MRN\_f1e3\_zeta1e16, yellow line) from the protostar formation.
Thus, the average velocity of the outflow head is found to be $\sim1.8\ \rm{km}\ \rm{s}^{-1}$ for model\_MRN\_f1e2\_zeta1e16 (orange), $\sim2.4\ \rm{km}\ \rm{s}^{-1}$ for model\_MRN\_f1e3\_zeta1e16 (yellow).
This result indicates that high cosmic ray ionization models have about twice the velocity of low cosmic ray ionization models.

Figure \ref{fig:outflow_mass} shows the time evolution of the outflow mass.
In all models, the outflow mass is monotonically increasing.
Outflow mass is inversely correlated with disk mass, suggesting that outflow activity is related to disk growth.
Models with a high cosmic ray ionization rate have larger masses than those with a low ionization rate.
Regarding the dust fraction, the low dust fraction model (yellow) has a more active outflow than the high dust fraction model (orange) with high cosmic ray ionization.
Thus, Figure \ref{fig:outflow_mass} suggests also that the dust fraction affects the evolution of outflow in a high cosmic ray ionization environment.

Figure \ref{fig:outflow_J} shows the time evolution of the outflow specific angular momentum.
The dust fraction has little effect on the specific angular momentum of the outflow.
If we compare the high and low cosmic-ray ionization, the difference among the models is about a factor of two at the end of the simulation.
This means that the outflow will more efficiently remove the angular momentum from the central region in high cosmic-ray ionization models.
Therefore, in models with a high cosmic ray ionization rate, the disk disappears at the end of the simulation.
This suggests that the high cosmic ray ionization environments have a more active outflow and that the angular momentum is removed from the disk.

\section{DISCUSSION}

\subsection{Impact on disk formation}
In all models considered in this paper, the circumstellar disk is formed at $t \sim 4.3\times10^4$ yr.
At the time of the formation, the centrifugal radius and mass of the disks were $2-4$ AU and $0.01-0.02 \Msun$, respectively.
It is also shown that disks can grow at low cosmic ray ionization rates ($\zeta_{\rm{CR}}=10^{-17}\rm{s^{-1}}$), but disappear at high cosmic ray ionization rates ($\zeta_{\rm{CR}}=10^{-16}\rm{s^{-1}}$). 
Figures \ref{fig:xy_density_map} and \ref{fig:t_J_comp} show a result that is consistent with the results of \citet{2020A&A...639A..86K} in which the formation of the disk strongly depends on the cosmic ray ionization rate.

Dust fraction, on the other hand, refers to the amount of dust grains in the cloud.
Dust grains are sinks for electrically charged particles.
Therefore, as the dust fraction decreases, the resistivity in the central region decreases (conductivity increases).
Therefore, it is expected that the removal of angular momentum by magnetic braking would be more efficient, suppressing the formation and evolution of disks in low dust fraction models ($Z=0.1\Zsun$).
However, our results did not show as pronounced a difference as the difference in cosmic ray ionization.
Thus, in the parameter range we considered, disk formation is more strongly affected by cosmic ray ionization than by dust fraction.
\subsection{Activity of outflow}
Outflow is one of the mechanisms that remove angular momentum from the central region.
In this simulation, outflows form in all models, even with different cosmic ray ionization rates and dust fractions.
We found that the nature of the outflow does differ depending on the cosmic ray ionization rate and dust fraction.
For example, the timing of outflow formation is earlier in models with a high cosmic ray ionization rate than in models with a low rate.
In the case of a high cosmic ray ionization rate, the mass and angular momentum of the outflow are larger for the low dust fraction model (model\_MRN\_f1e3\_zeta1e16) than for the high dust fraction model (model\_MRN\_f1e2\_zeta1e16).
On the other hand, within the low cosmic ray ionization models, there is little difference depending on the dust fraction.

\begin{figure}
    \centering
    \includegraphics[width=0.65\linewidth,angle=-90]{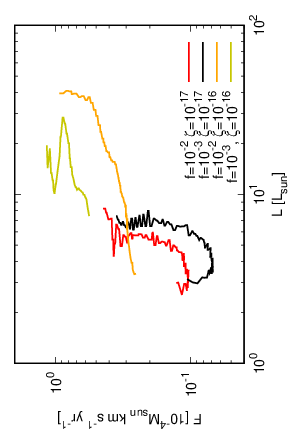}
    \caption{Outflow force $F$ versus luminosity.
    The luminosity is calculated with eq.3 in \citet{2012MNRAS.427.1182S}.
    The red, black, orange, and yellow lines show the results of model\_MRN\_f1e2\_zeta1e17, model\_MRN\_f1e3\_zeta1e17, model\_MRN\_f1e2\_zeta1e16, and model\_MRN\_f1e3\_zeta1e16, respectively.
    The arrows show the direction of time evolution.}
    \label{fig:F_outflow}
\end{figure}

Figure \ref{fig:F_outflow} plots the momentum $F$ per unit time injected into the outflow versus the luminosity of the central source $L$ \citep[which can be compared with figure 7 ]{2004A&A...426..503W}.

Here,
\begin{eqnarray}
    F = \frac{M_{outflow}V_{outflow}}{\Delta t},\\
    L = f G\frac{\dot{M}_{star}M_{star}}{3R_\odot},
\end{eqnarray}
where $\Delta t$ is the time after the launching of the outflow and $f$ is the ratio that the gravitational potential converts to radiational energy, and $f = 0.25$.
In high cosmic ray ionization models (orange and yellow lines), $F$ is larger than in low models (red and black lines), and the luminosity rapidly increases at the end of simulations.
This rapidly increasing luminosity is due to the disk gas rapidly falling on the central object.
While low cosmic ray ionization models do not have rapidly increasing luminosity at the end of simulations, they do have rapidly decreasing luminosity.
However, in low cosmic ray ionization models, outflow momentum rapidly increases at the end of the simulation.

From Figure ~\ref{fig:F_outflow}, we suggest that outflow momentum differs depending on the environment, such as cosmic ray ionization rate and dust fraction.
For instance, in Figure 7 of \citet{2004A&A...426..503W}, the plots are scattered not only along the horizontal axis (luminosity) but also along the vertical axis (outflow momentum per unit time).
Horizontal scattering is thought to be due to the time evolution of the objects.
However, it is not clear whether vertical scattering is due to the evolution of objects or not.
In figure ~\ref{fig:F_outflow}, there is a clear difference among the models.
For instance, models with a high cosmic ray ionization rate have greater outflow momentum than ones with a low cosmic ray ionization rate.
Thus, the environment of a star-forming region may affect the outflow momentum through the difference in ionization.

\section{SUMMARY}
In this paper, we study the formation and early evolution of young stellar objects (YSOs) using three-dimensional non-ideal magnetohydrodynamic (MHD) simulations to investigate the effect of cosmic ray ionization rate and dust fraction (or amount of dust grains) on circumstellar disk formation.
Our results are summarized as follows:
\begin{enumerate}
    \item 
        Circumstellar disks are formed in all simulations at $t\sim4.3\times10^4$ yr.
        Disks eventually grow up $r\sim10$ AU in low cosmic ray ionization models (model\_MRN\_f1e3\_zeta1e17 and model\_MRN\_f1e2\_zeta1e17), however, they disappear in high cosmic ray ionization models (model\_MRN\_f1e3\_zeta1e16 and model\_MRN\_f1e2\_zeta1e16) at the end of simulations.
        While the impact on disk formation and evolution of the dust fraction is smaller than the cosmic ray ionization rate.
    \item
        All simulations were outflow driven.
        In high cosmic ray ionization and low dust fraction models, the outflow has a greater angular momentum and outflow force.
        Thus, the cosmic ray ionization rate and dust fraction can influence outflow formation and evolution. 
\end{enumerate}

\section*{Acknowledgements}
We thank the anonymous referee for her/his insightful comments.
A Cray XC50 at the Center for Computational Astrophysics at the National Astronomical Observatory of Japan was used to calculate the simulation.

\section*{Data Availability}

The simulation data will be shared on reasonable request with the corresponding author.



\bibliographystyle{mnras}
\bibliography{export-bibtex}







\bsp	
\label{lastpage}
\end{document}